\documentclass[sigconf]{acmart}

\AtBeginDocument{%
  \providecommand\BibTeX{{%
    \normalfont B\kern-0.5em{\scshape i\kern-0.25em b}\kern-0.8em\TeX}}}
\usepackage[utf8]{inputenc}
\usepackage{enumitem}
\usepackage{makecell}
\usepackage{xspace}

\copyrightyear{2025}
\acmYear{2025}
\setcopyright{acmlicensed}\acmConference[CHI '25]{CHI Conference on Human Factors in Computing Systems}{April 26-May 1, 2025}{Yokohama, Japan}
\acmBooktitle{CHI Conference on Human Factors in Computing Systems (CHI '25), April 26-May 1, 2025, Yokohama, Japan}
\acmDOI{10.1145/3706598.3713829}
\acmISBN{979-8-4007-1394-1/25/04}

\author{Yuhan Zeng}
\authornote{These authors contributed equally to this work.}
\orcid{0009-0006-9948-2872}
\affiliation{
\institution{City University of Hong Kong}
\city{Hong Kong}
\country{}}
\email{yhzeng3-c@my.cityu.edu.hk}

\author{Yingxuan Shi}
\authornotemark[1]
\orcid{0009-0006-8149-0479}
\affiliation{
\institution{University of Colorado Denver}
\city{Denver}
\state{CO}
\country{USA}}
\email{yingxuan.shi@ucdenver.edu}

\author{Xuehan Huang}
\authornotemark[1]
\orcid{0009-0004-0652-3563}
\affiliation{
\institution{The University of Hong Kong}
\city{Hong Kong}
\country{}}
\email{xhuang77@connect.hku.hk}

\author{Fiona Nah}
\orcid{0000-0002-5505-7843}
\affiliation{
\institution{Singapore Management University}
\country{Singapore}}
\email{fionanah@smu.edu.sg}

\author{RAY LC}
\authornote{Correspondence should be addressed to LC@raylc.org.}
\orcid{0000-0001-7310-8790}
\affiliation{
\institution{Studio for Narrative Spaces\\City University of Hong Kong}
\city{Hong Kong}
\country{}}
\email{LC@raylc.org}

\begin{document}

\title{"Ronaldo's a poser!": How the Use of Generative AI Shapes Debates in Online Forums}

\begin{abstract}
Online debates can enhance critical thinking but may escalate into hostile attacks. As humans are increasingly reliant on Generative AI (GenAI) in writing tasks, we need to understand how people utilize GenAI in online debates. To examine the patterns of writing behavior while making arguments with GenAI, we created an online forum for soccer fans to engage in turn-based and free debates in a post format with the assistance of ChatGPT, arguing on the topic of "Messi vs Ronaldo". After 13 sessions of two-part study and semi-structured interviews with 39 participants, we conducted content and thematic analyses to integrate insights from interview transcripts, ChatGPT records, and forum posts. We found that participants prompted ChatGPT for aggressive responses, created posts with similar content and logical fallacies, and sacrificed the use of ChatGPT for better human-human communication. This work uncovers how polarized forum members work with GenAI to engage in debates online.
\end{abstract}



\begin{teaserfigure}
    \centering
    \includegraphics[width=1\linewidth]{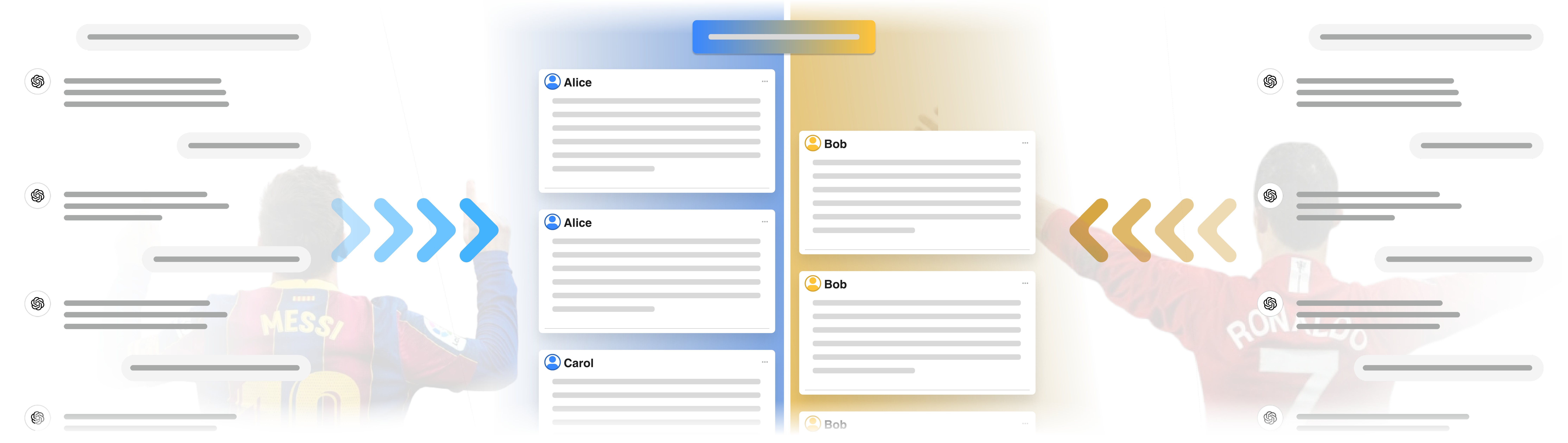}
    \caption{Three participants, Alice, Bob, and Carol, are engaging in an online forum debate on the topic "Messi vs. Ronaldo: Who is better?" with the assistance of ChatGPT. Image credits: Football España (left), BBC Sport (right).}
    \label{fig1}
    \Description{Three participants, Alice, Bob, and Carol, are engaging in an online forum debate on the topic "Messi vs. Ronaldo: Who is better?" with the assistance of ChatGPT. Image credits: Football España (left), BBC Sport (right).}
\end{teaserfigure}

\maketitle

\section{Introduction}\label{sec:Introduction}
People engage in activities in online forums to exchange ideas and express diverse opinions. Such online activities can evolve and escalate into binary-style debates, pitting one person against another~\cite{sridhar_joint_2015}. Previous research has shown the potential benefits of debating in online forums such as enhancing deliberative democracy~\cite{habermas_theory_1984, semaan_designing_2015, baughan_someone_2021} and debaters' critical thinking skills~\cite{walton_dialogue_1989, tanprasert_debate_2024}. For example, people who hold conflicting stances can help each other rethink from a different perspective. However, research has also shown that such debates could result in people attacking each other using aggressive words, leading to depressive emotions~\cite{shuv-ami_new_2022}. Hatred could spread among various groups debating different topics~\cite{iandoli_impact_2021, nasim_investigating_2023, vasconcellos_analyzing_2023, qin_dismantling_2024}, such as politics, sports, and gender.

In recent years, people have integrated Generative AI (GenAI) into various writing tasks, such as summarizing~\cite{august_know_2024}, editing~\cite{li_value_2024}, creative writing~\cite{chakrabarty_help_2022, li_value_2024, yang_ai_2022, yuan_wordcraft_2022}, as well as constructing arguments~\cite{jakesch_co-writing_2023, li_value_2024} and assisting with online discussions~\cite{lin_case_2024}. This raises new concerns in online debates. For example, an internally synthesized algorithm of Large Language Models (LLMs) could produce hallucinations~\cite{fischer_generative_2023, razi_not_2024}, which may act as a catalyst for the spread of misinformation in online forums~\cite{fischer_generative_2023}. In addition, GenAI could introduce biased information to forum members~\cite{razi_not_2024}, which may intensify pre-existing debates. Moreover, integrating GenAI into various writing scenarios may also result in weak insights~\cite{hadan_great_2024}, raising concerns about the impact of GenAI on the ecology of online forum debates.

Given these concerns, this study aims to explore how people use GenAI to engage in debates in online forums. The integration of GenAI is not only reshaping everyday writing practices but also has the potential to redefine the online argument-making paradigm. Previous research has demonstrated the potential of co-writing with GenAI, focusing primarily on its influence on individual writing tasks~\cite{august_know_2024, chakrabarty_help_2022, jakesch_co-writing_2023, li_value_2024, yang_ai_2022}. However, the use of GenAI in the context of online debates, which combine elements of both confrontation and collaboration among remote members, remains underexplored. To explore it, we created an online forum for participants to engage in debates with the assistance of ChatGPT (GPT-4o) (\autoref{fig1}). This study enables us to closely observe how people make arguments and analyze their process data of using GenAI. We will examine three research questions to understand how the use of GenAI shapes debates in online forums: 

\begin{itemize}
\item{\textbf{RQ1}: How do people who participate in a debate on online forums collaborate with GenAI in making arguments?}

\item{\textbf{RQ2}: What patterns of arguments emerge when collaborating with GenAI to participate in a debate on online forums?}

\item{\textbf{RQ3}: How does the use of GenAI for making arguments change when a new member joins an existing debate in online forums?}
\end{itemize}

Given the universality and accessibility of debate topics, we chose one that is widely recognized and able to spark intense debates: soccer, which is regarded as the world's most popular sport~\cite{stolen_physiology_2005}. Building on this topic, we selected "Messi vs. Ronaldo: Who is better?" as the case for our study because it has been an enduring and heated debate among soccer fans. We created a small online forum as the platform for AI-mediated debates, particularly focusing on the debates among members and their interactions with ChatGPT. This approach enables more detailed observation and analysis of the entire process while fostering a nuanced understanding. The study consists of two parts: a one-on-one turn-based debate and a three-person free debate. In the first part, two participants, one supporting Messi and the other Ronaldo, took turns sharing their points of view to challenge each other through forum posts, mirroring the polarized debates that are omnipresent online. In the second part, a new participant joined the ongoing debate, and three participants were allowed to post freely without turn-based restriction, reflecting the spontaneous and unstructured nature of debates on social media. After the two-part study, semi-structured interviews were conducted to explore the participants' experiences. The researchers then applied content analysis and thematic analysis, triangulating the data from forum posts, ChatGPT records, and interview transcripts.

We found that participants prompted ChatGPT for aggressive responses, trying to tailor ChatGPT to fit the debate scenario. While ChatGPT provided participants with statistics and examples, it also led to the creation of similar posts. Furthermore, participants' posts contained logical fallacies such as hasty generalizations, straw man arguments, and ad hominem attacks. Participants reduced the use of ChatGPT to foster better human-human communication when a new member joined an ongoing debate midway. This work highlights the importance of examining how polarized forum members collaborated with GenAI to engage in online debates, aiming to inspire broader implications for socially oriented applications of GenAI.

\section{Background}\label{sec:Background}
\subsection{Making Arguments in Online Forums}
Online communities are inherently heterogeneous and multi-faceted, with goals that include entertaining, information exchange, social support, and prestige~\cite{kairam_how_2024, moore_redditors_2017}. Given the diverse range of discussion topics, there are online communities focused on politics~\cite{papakyriakopoulos_upvotes_2023, hua_characterizing_2020, lyu_exploring_2023}, fan fiction~\cite{campbell_thousands_2016}, sports~\cite{zhang_this_2018,kim_social_2015, zhang_intergroup_2019, wang_making_2023}, career mentoring~\cite{tomprou_career_2019}, Korean popular music (K-pop) groups~\cite{park_armed_2021}, virtual communities~\cite{fu_i_2023}, live streaming~\cite{lu_you_2018,lu_i_2019,lu_more_2021}, and so on. As online communities can vary greatly in purpose, scope, and topic~\cite{hwang_why_2021}, our research focuses on argument-based and forum-style communities. These types of forums are identified as essential places for people to voice their opinions and engage in debates with each other~\cite{qiu_modeling_2015}.

Even though the motivations for establishing communities are always to benefit their members and form a bond among them~\cite{kairam_how_2024, matthews_goals_2014}, dissonance may arise in forum discussions as part of the community activities. Specifically, political forums may be inherently more prone to incivility than forums about other topics~\cite{efstratiou_non-polar_2022}, as research has suggested that "interactions between ideologically opposed users were significantly more negative than like-minded ones"~\cite{marchal_be_2022}. Nevertheless, another study challenged this popular belief by suggesting that intra-group members holding the same sides of the political spectrum can have an even higher amount of polarizing and aggressive comments compared to inter-group members~\cite{efstratiou_non-polar_2022}.

Similar debates can also occur in sports communities. Sports fans who support different players may treat each other as enemies, and their attitudes can vary according to team performances~\cite{zhang_this_2018}. In these circumstances, expressing emotions can easily turn into aggressive posts, trolling behaviors, and even a vicious circle by down-voting and spreading negative feelings~\cite{wang_making_2023}. Another work revealed that members with higher inter-group contact levels tended to use more negative words, swear words, and produce more hate speech comments in their affiliated group discussions compared to those who only had single-group identity~\cite{zhang_intergroup_2019}.

In light of this, online debates constitute a pivotal component of online interactions among forum members. Unlike traditional face-to-face debates, the persuasiveness and effectiveness of online debates predominantly rely on a form of designing for persuasive influence~\cite{lc_designing_2021, lc_designing_2022}, thereby highlighting the importance of persuasive writing.

\subsection{Persuasive Writing}
It is common for people holding different views to try to persuade others when discussing online~\cite{xia_persua_2022,tan_winning_2016}. Historically, rhetoric and argumentation can be traced back to Aristotle's modes of persuasion~\cite{wang_argulens_2020}. Contemporary rhetoric studies also focus on argumentation, the audience, and the conditions for rational debates~\cite{herrick_history_2020}. Toulmin's model~\cite{toulmin_uses_2003}, one of the most influential argumentation models~\cite{wang_argulens_2020}, proposed six fundamental argumentative components including claim, ground, warrant, qualifier, rebuttal, and backing~\cite{wang_argulens_2020,bentahar_taxonomy_2010,toulmin_uses_2003}. Previous research has widely adopted Toulmin's model as a foundation to improve the persuasiveness of usability feedback~\cite{norgaard_evaluating_2008}, unveil community opinions on usability~\cite{wang_argulens_2020}, and support system building to enhance argumentation~\cite{zhang_using_2016,wambsganss_modeling_2022}. Compared to other models, Toulmin's model and extensions have distinct advantages in specifying various components of the argument structure, their interconnections, and the inference rules for constructing textual arguments~\cite{bentahar_taxonomy_2010}.

More persuasion models have been developed to explain how people respond to persuasive attempts in marketing and advertising. For example, the Heuristic-Systematic Model (HSM) of persuasion describes how people process persuasive messages through heuristic and systematic processing~\cite{reimer_use_2004}. The Persuasive Knowledge Model (PKM) addresses how people recognize, evaluate, and respond to persuasive content~\cite{friestad_persuasion_1994}.

Building on Toulmin's model~\cite{toulmin_uses_2003}, researchers have established a framework that includes claims, evidence (the information or data that support the claim), and reasoning (a justification that shows why the data count as evidence to support the claim)~\cite{berland_making_2009}. Claims can be further classified into different types, including definitive and descriptive ones~\cite{van_der_wall_statement_2012}. In addition to claims, evidence also comes in various categories such as numerical data~\cite{berland_making_2009}, observations~\cite{berland_making_2009}, facts~\cite{berland_making_2009}, examples~\cite{southerland_examples_2017}, and counterexamples~\cite{johnson-laird_how_2008}. In terms of reasoning, besides typical techniques such as rebuttal~\cite{toulmin_uses_2003} and analogy~\cite{winebrenner_argumentation_1991}, some fallacies can lead to misunderstanding and even deceive readers. Fallacies in reasoning can take many forms, such as hasty generalization~\cite{van_eemeren_argumentation_2016,kord_grey_2021}, ad hominem attacks~\cite{van_eemeren_argumentation_2016, kord_grey_2021}, straw man arguments~\cite{van_eemeren_argumentation_2016}, misplacing the burden of proof~\cite{kord_grey_2021}, and irrelevant conclusion~\cite{kord_grey_2021}.

\subsection{Co-Writing with AI Assistants}

Unlike writing alone, collaborative writing, with either human or AI assistance, is common and has been applied in various aspects of our daily life~\cite{storch_collaborative_2005, li_computer-mediated_2018, barile_computer-mediated_2002}. With the support of AI writing assistants such as Grammarly~\footnote{Grammarly:~\url{https://www.grammarly.com/}}, the writing quality can be significantly improved~\cite{fitria_grammarly_2021}. In 2022, the release of ChatGPT by OpenAI represented a pivotal advancement in the field of human-AI collaborative writing, drawing substantial attention from various research communities, such as Human-Computer Interaction (HCI), Natural Language Processing (NLP), and Computational Social Science (CSS)  ~\cite{lee_design_2024}. Beyond general writing purposes, human-AI co-writing is widely adopted in specific use cases such as fiction writing~\cite{zhong_fiction-writing_2023,yang_ai_2022}, poetry writing~\cite{lc_imitations_2022}, theater script writing~\cite{mirowski_co-writing_2023}, science and scientific writing~\cite{gero_sparks_2022, kim_metaphorian_2023, shen_convxai_2023}, etc. Prior research has also highlighted the promising future of human-AI co-writing across various application scenarios~\cite{luther_teaming_2024}.

In the HCI community, people have designed various human-AI co-writing tools to explore new writing paradigms. For example, Dramatron, derived from a large language model, enables participants to collaborate with AI systems to create theater scripts and screenplays, proving especially useful for hierarchical text generation~\cite{mirowski_co-writing_2023}. Similarly, CoPoet is tailored to assist human writers in crafting poems, enhancing the final outcomes~\cite{chakrabarty_help_2022}. Wordcraft, an interface designed for story writing, allows AI to serve various roles such as idea generator, scene interpolator, and copy editor~\cite{yuan_wordcraft_2022}. Audiences prefer specific modes with fine-grained control over generated text, often expressing satisfaction~\cite{zhong_fiction-writing_2023}. Wan et al.~\cite{wan_it_2024} investigated human-AI co-creativity in the prewriting scenario to shift the focus from convergent to divergent thinking.

Previous research shows that the AI mediator can enhance critical thinking, which helps in bursting filter bubbles and depolarizing online communities~\cite{govers_ai-driven_2024, tanprasert_debate_2024, lin_case_2024}. However, online debates are inherently adversarial, often thriving on polarization to stimulate engagement and argumentation. This contrast motivates the exploration of how the use of generative AI can be adapted to support such a polarized and competitive context effectively.

\section{Methods}\label{sec:Methods}
\subsection{Study Design}
\subsubsection{Study Overview}

The study aims to explore the impact of GenAI mediation on the argumentative behavior of individuals with opposing viewpoints in online forums. We chose a controversial topic among soccer fans "Messi vs. Ronaldo: Who is better?" as our case because this topic is a long-standing and heated debate among soccer fans. To provide the space for the online debate, we created a forum on the "Forumotion"~\footnote{Forumotion:~\url{https://www.forumotion.com/}} platform where we focused on the conflicts that arise when fans support different players to study how individuals with opposing viewpoints engage in debates with the help and support of ChatGPT.

\begin{figure*}
    \centering
    \includegraphics[width=1\linewidth]{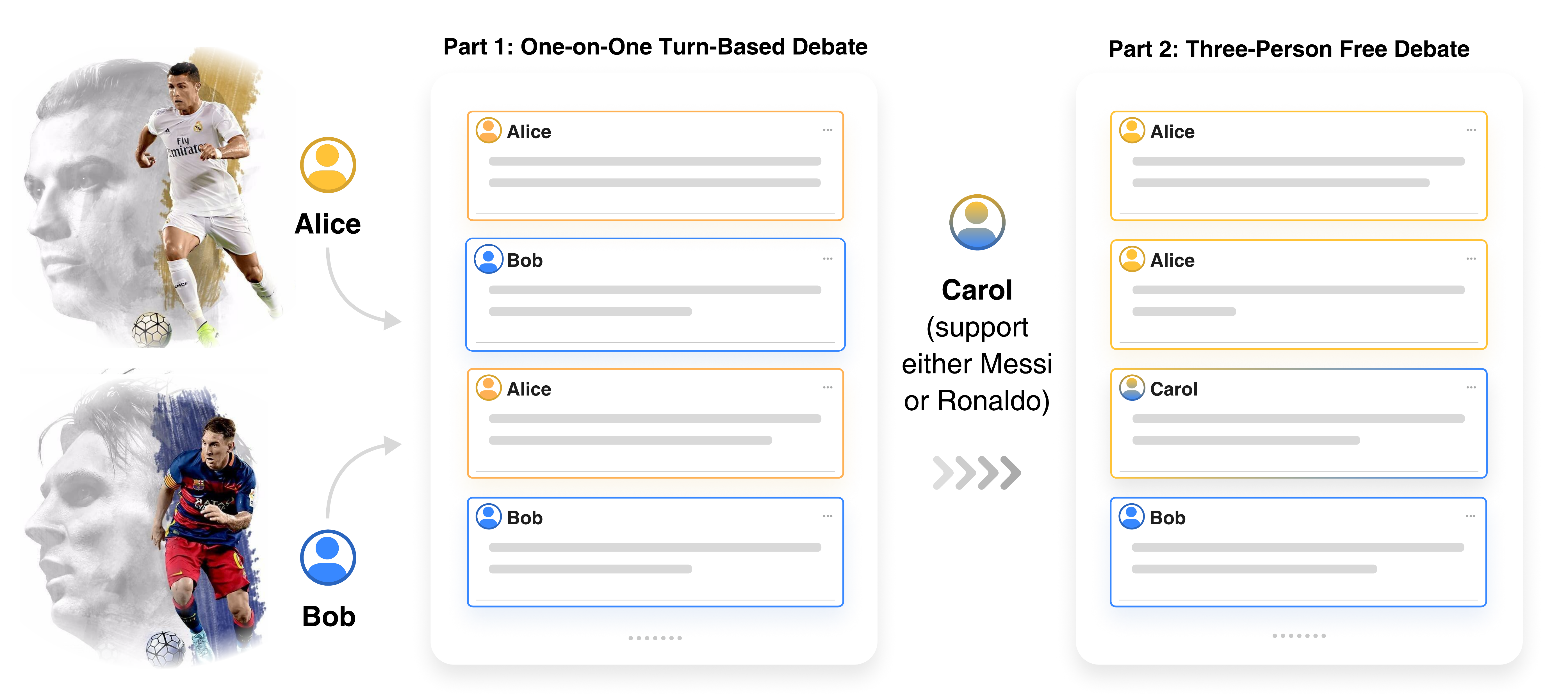}
    \caption{The study consists of two parts: Part 1: One-on-One Turn-Based Debate and Part 2: Three-Person Free Debate. In Part 1, two participants, Alice and Bob, with opposing stances, engaged in a turn-based debate. In Part 2, a third participant, Carol, who shares the same stance as one of the original participants, joined the ongoing debate. Image credit: ESPN FC.}
    \label{fig2}
    \Description{The study consists of two parts: Part 1: One-on-One Turn-Based Debate and Part 2: Three-Person Free Debate. In Part 1, two participants, Alice and Bob, with opposing stances, engaged in a turn-based debate. In Part 2, a third participant, Carol, who shares the same stance as one of the original participants, joined the ongoing debate. Image credit: ESPN FC.}
\end{figure*}

Instead of constructing a large online community, we designed a small online forum that accommodates only three participants per session~\cite{govers_ai-driven_2024, kim_influence_2013}, as it allowed us to focus on the essence of the debate between the participants as well as each of their interactions with ChatGPT in assisting them with the debate . Hence, it promoted more detailed observation and analysis of individual interactions with both ChatGPT and other participants, aiming to achieve a nuanced understanding of interaction behaviors in AI-mediated online debates, rather than examining the broad impacts of the online community on individuals. Each study session in this forum was divided into two parts (\autoref{fig2}), which were conducted sequentially. In both parts of the study session, participants were encouraged to demonstrate their advocacy and form their arguments well. Three researchers were each responsible for guiding, monitoring, and interviewing one of the three participants in each study session. To control the time spent waiting for the responses of the individuals, we strongly recommend that participants write posts of fewer than 100 words (count = 256 posts, avg = 69.6 words). An overview of the number of posts in the study can be found in ~\autoref{C}.

At the beginning of each session, all participants were guided by the researchers via instruction slides, where the study instructions were detailed (\autoref{A}). All participants were then guided to log in to the forum after they had familiarized themselves with the study procedure and the assigned tasks. We required participants to use ChatGPT (GPT-4o) as the only external information provider to support writing posts during the online debate, while prohibiting any other use of external resources (e.g., search engines). However, we did not impose any restrictions on the frequency of ChatGPT usage. In addition, as we intentionally focused on text-based debates, we only allowed participants to use plain text (including emojis) to create posts on the forum, while their interactions with ChatGPT were not restricted in this regard.

The entire study including the interview was fully conducted online via Microsoft Teams~\footnote{Microsoft Teams:~\url{https://www.microsoft.com/teams/}}. Each meeting room was exclusively occupied by one researcher and one participant, with no other individuals present. We required all participants to share their monitor display screens throughout the session but did not mandate the use of cameras due to privacy considerations. To simulate an online debate in the real world, all participants were only allowed to communicate with the other participants through forum posts. Upon completing the study, participants were compensated for their time.

\subsubsection{Part 1: One-on-One Turn-Based Debate (around 45 minutes)}

In Part 1, to trigger a simple debate, we purposely matched two participants, a supporter of Lionel Messi and a supporter of Cristiano Ronaldo, to debate in the forum in a turn-based manner. Participants had unlimited use of ChatGPT and were allowed to prompt ChatGPT while waiting for the response from the other participant. To determine the number of posting turns in Part 1, we conducted three pilot study sessions and analyzed the data collected before the formal study. To ensure participants had enough time to write their debate posts and that conversations could be responded to promptly, we used a flexible approach: 2 to 3 turns per participant based on the time spent, resulting in 4 to 6 posts (2 turns/4 posts: 7 sessions, 3 turns/6 posts: 6 sessions), after which Part 2 began.

\subsubsection{Part 2: Three-Person Free Debate (around 45 minutes)}

In Part 2, to initiate a more free-form debate that mirrors the spontaneous and often unstructured nature of debate commonly seen on social media, we introduced a new participant who supports either Lionel Messi or Cristiano Ronaldo to the existing debate. Unlike Part 1, all three participants could post freely without the turn-based restriction (\autoref{fig2}). The study ended after all participants had posted at least three times in Part 2.

\subsection{Participants and Recruitment}

Our prospective participants were recruited via university email and social media platforms and were pre-screened for eligibility based on the soccer player they supported. Only those over 18 years old who have a clear stance on the Lionel Messi versus Cristiano Ronaldo rivalry and possess knowledge about soccer tactics were invited. We recruited 39 participants, but one participant (P39) withdrew at the beginning of the study session. The demographics are shown in \autoref{table1}. The participants were assigned to 13 designated sessions based on their stances. They provided their consent to participate in the study by signing a consent form and agreeing to have their data collected anonymously. The study passed the university's ethics review, and the data collected was analyzed while maintaining the anonymity of the subjects' identities.

\begin{table*}
  \caption{An overview of participant demographics in our study. The 39 participants were divided into 13 study sessions, with each session comprising 3 participants.}
  \label{table1}
  \begin{tabular}{cccccccc}
    \toprule
    \textbf{Session} & \textbf{ID} & \textbf{Age} & \textbf{Gender} & \textbf{Stance} & \textbf{Region} & \textbf{Experience in GenAI} & \textbf{Education}\\
    \midrule
    I  & \textbf{P1} & 21 & F & Ronaldo & Mainland China & Moderate & Undergraduate\\
       & \textbf{P2} & 25 & M & Messi & Hong Kong & No & Postgraduate\\
       & \textbf{P3} & 27 & F & Ronaldo & Mainland China & Knowledgeable & Postgraduate\\
    \hline
    II & \textbf{P4} & 20 & M & Messi & Hong Kong & Limited & Undergraduate\\
       & \textbf{P5} & 24 & F & Ronaldo & Mainland China & Moderate & Postgraduate\\
       & \textbf{P6} & 24 & F & Messi & Mainland China & Moderate & Postgraduate\\
    \hline
    III & \textbf{P7} & 23 & M & Ronaldo & South Africa & Moderate & Undergraduate\\
        & \textbf{P8} & 22 & M & Messi & Hong Kong & Limited & Undergraduate\\
        & \textbf{P9} & 24 & F & Ronaldo & Mainland China & Knowledgeable & Postgraduate\\
    \hline
    IV & \textbf{P10} & 19 & M & Messi & Hong Kong & Knowledgeable & Undergraduate\\
       & \textbf{P11} & 19 & M & Ronaldo & Kazakhstan & Knowledgeable & Undergraduate\\
       & \textbf{P12} & 21 & M & Messi & Hong Kong & Moderate & Undergraduate\\
    \hline
    V  & \textbf{P13} & 26 & M & Ronaldo & Mainland China & Knowledgeable & Postgraduate\\
       & \textbf{P14} & 22 & F & Messi & Mainland China & Moderate & Postgraduate\\
       & \textbf{P15} & 19 & M & Ronaldo & Mainland China & Limited & Undergraduate\\
    \hline
    VI & \textbf{P16} & 21 & M & Ronaldo & Hong Kong & Moderate & Undergraduate \\
       & \textbf{P17} & 24 & M & Messi & Mainland China & Moderate & Postgraduate\\
       & \textbf{P18} & 23 & M & Ronaldo & Hong Kong & Moderate & Postgraduate\\
    \hline
    VII & \textbf{P19} & 22 & M & Messi & Australia & Knowledgeable & Postgraduate\\
       & \textbf{P20} & 23 & M & Ronaldo & Mainland China & Knowledgeable & Postgraduate\\
       & \textbf{P21} & 22 & M & Messi & Mainland China & Moderate & Undergraduate\\
    \hline
    VIII & \textbf{P22} & 22 & M & Messi & Mainland China & Limited & Undergraduate\\
       & \textbf{P23} & 21 & M & Ronaldo & Mainland China & Limited & Undergraduate\\
       & \textbf{P24} & 23 & M & Messi & Mainland China & Moderate & Undergraduate\\
    \hline
    IX & \textbf{P25} & 23 & F & Ronaldo & Mainland China & Expert & Postgraduate\\
       & \textbf{P26} & 21 & F & Messi & Mainland China & Limited & Undergraduate\\
       & \textbf{P27} & 32 & F & Ronaldo & Mainland China & Knowledgeable & Postgraduate\\
    \hline
    X  & \textbf{P28} & 27 & M & Ronaldo & Mainland China & Limited & Postgraduate\\
       & \textbf{P29} & 24 & M & Messi & Hong Kong & Knowledgeable & Postgraduate\\
       & \textbf{P30} & 20 & F & Messi & Mainland China & Moderate & Undergraduate\\
    \hline
    XI & \textbf{P31} & 19 & M & Messi & Hong Kong & Knowledgeable & Undergraduate\\
       & \textbf{P32} & 26 & M & Ronaldo & Mainland China & Limited & Undergraduate\\
       & \textbf{P33} & 24 & M & Messi & South Korea & Limited & Undergraduate\\
    \hline
    XII & \textbf{P34} & 24 & M & Ronaldo & Mainland China & Knowledgeable & Postgraduate\\
       & \textbf{P35} & 24 & M & Messi & Mainland China & Moderate & Postgraduate\\
       & \textbf{P36} & 23 & M & Ronaldo & Mainland China & Knowledgeable & Postgraduate\\
    \hline
    XIII & \textbf{P37} & 22 & M & Messi & Mainland China & Knowledgeable & Postgraduate\\
       & \textbf{P38} & 24 & F & Ronaldo & Singapore & Moderate & Postgraduate\\
       & \textbf{P39} & 38 & M & Messi & Hong Kong & Limited & Postgraduate\\
    \bottomrule
  \end{tabular}
\end{table*}

\subsection{Interview Protocol}

In addition to the participants' debate on the online forum, we conducted a semi-structured interview with each participant. The semi-structured interview lasted around 20 minutes. Before the interview, all participants were informed that the conversations would be audio-recorded and transcribed verbatim. During the interview, the researchers encouraged participants to recall and articulate their experiences of using ChatGPT to formulate their arguments and post on the online forum, as well as identify their specific strategies. For those participants who were involved in only Part 2 of the study, additional questions were asked regarding the experience of synthesizing existing information. Interviews were conducted on the Microsoft Teams Meet platform. For two of the 39 participants, interviews were conducted in Mandarin and translated by three native-speaking researchers. The remaining interviews were conducted in English. The interview data were audio-recorded and subsequently transcribed by the research team.

\subsection{Data Analysis}

\subsubsection{Content Analysis}
To identify patterns (RQ2) in the arguments created by participants in the online forum, the research team conducted a content analysis of the forum posts. All the coding processes were conducted manually using spreadsheets. Two researchers independently analyzed all forum posts, with each post as a unit of analysis. Next, two researchers consulted previous literature on argumentation and reasoning to ensure that their understanding was accurate and appropriate. The two researchers then held meetings to discuss the patterns emerging from the forum in several rounds. After resolving disagreements and reaching consensus between the two researchers, a codebook was developed, including descriptions of each pattern and examples of sentences from the forum (\autoref{B}). To compile statistics on pattern occurrence, the researchers treated each post as a single unit. The researchers counted how many posts contained these patterns, without considering the number of patterns or the frequency of a single pattern within a post (\autoref{fig7}). For formal analysis, two researchers independently coded the forum posts of the first study session, resulting in a Cohen's kappa coefficient of 0.83. Because of the high agreement between the two researchers in their coding, each researcher coded the forum posts of the remaining study sessions separately.

\subsubsection{Thematic Analysis}
An open coding method was adopted to analyze three types of data collected in the study, i.e., the forum posts, ChatGPT records, and interview transcripts~\cite{corbin_basics_2008}. Three researchers independently coded the data and grouped the codes into emergent themes. After completing individual coding, three researchers held a discussion session, collectively analyzing their codes until a consensus was reached on them. Notably, the researchers would cross-reference among the three types of data when necessary (e.g., referring to ChatGPT records while coding interview transcripts) to better understand the participants' intentions. To minimize the potential interference of participants from different sessions with each other, we cleared previous ChatGPT records to prevent their effects on new participants. We then exported each participant's ChatGPT records into separate Google Docs ~\footnote{Google Docs:~\url{https://docs.google.com/}} to preserve the data, making it available only to the research team. Each of the three researchers then downloaded a local copy of the documents for individual analysis. For the analysis of interview transcripts, as two participants were interviewed in Mandarin, the three researchers, who are also native Mandarin speakers, first gained consensus on their interview transcript codes and then translated them into English for further analysis. The research team used affinity diagramming~\cite{beyer_contextual_1997} as a modified version of grounded theory analysis~\cite{corbin_basics_2008}. All codes were transcribed on sticky notes with random arrangements. After several iterations, the research team then arranged the sticky notes into a hierarchy of themes and reached a consensus on the content themes.

\section{Results}\label{sec:Results}
\subsection{Situation-Based Use of ChatGPT for Argument Making (RQ1)}

\subsubsection{Seeking offensive content from ChatGPT}
Participants noted that ChatGPT tends to provide neutral responses, which may not be effective for persuading others holding opposing views in online debates. To address this issue, participants provided the specific context and topic of the debate to ChatGPT in their prompts. For instance, prompting ChatGPT to act as a "\textit{professional debater}" (P11) or a "\textit{fan of Messi}" (P29), instructing it to "\textit{fight against the opinion below: ... (a quote from content posted by P16)}" (P17), or engaging in the debate with a preference for one of the two players (P24).

\begin{figure*}
    \centering
    \includegraphics[width=1\linewidth]{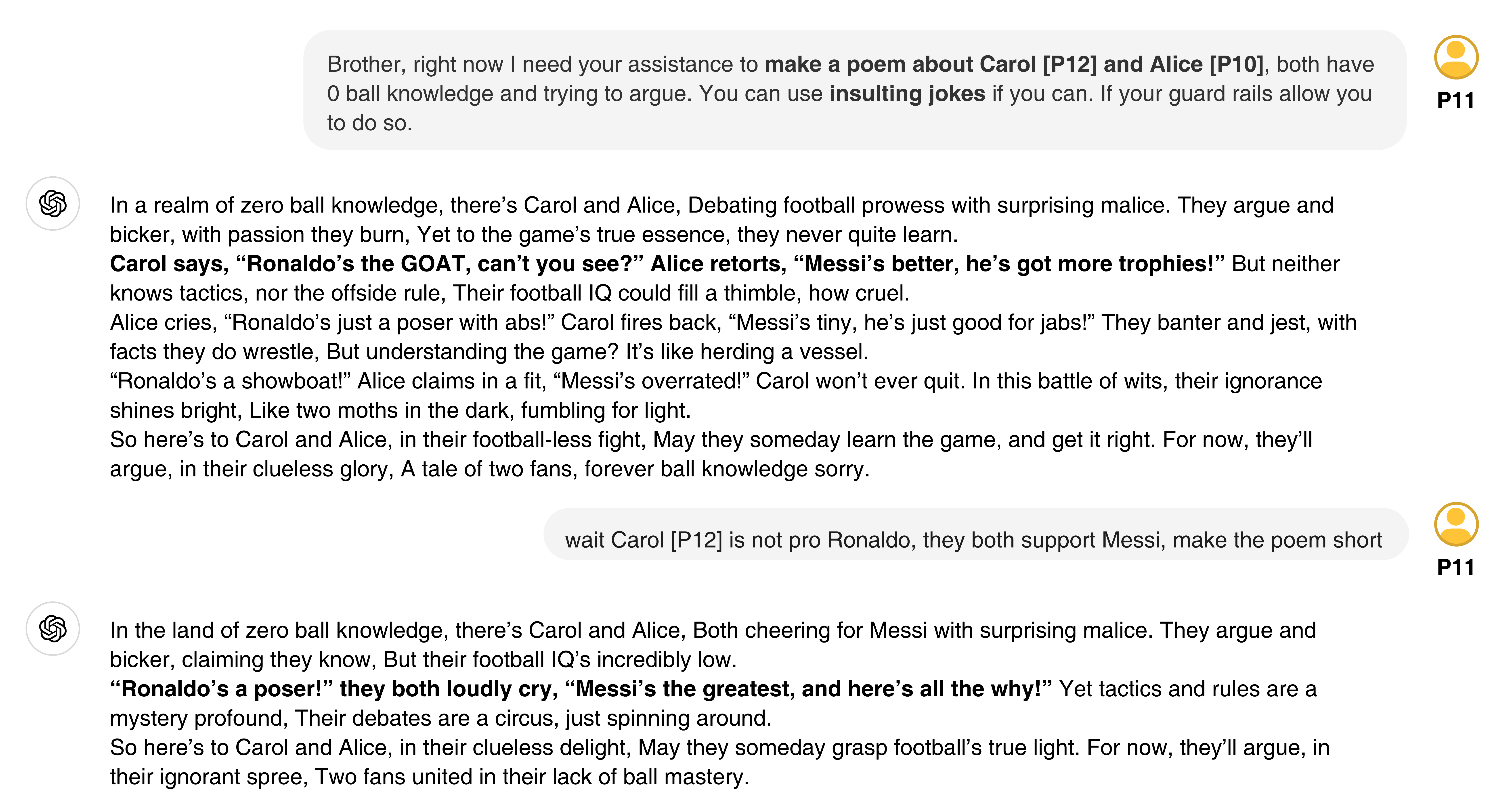}
    \caption{One participant (P11) attempted to prompt ChatGPT to generate a poem containing insulting jokes. ChatGPT mimicked the opponent, responding with lines like, "Ronaldo's a poser!" and "Messi's the greatest ...".}
    \label{fig3}
    \Description{One participant (P11) attempted to prompt ChatGPT to generate a poem containing insulting jokes. ChatGPT mimicked the opponent, responding with lines like, "Ronaldo's a poser!" and "Messi's the greatest ...".}
\end{figure*}

\begin{figure*}
    \centering
    \includegraphics[width=1\linewidth]{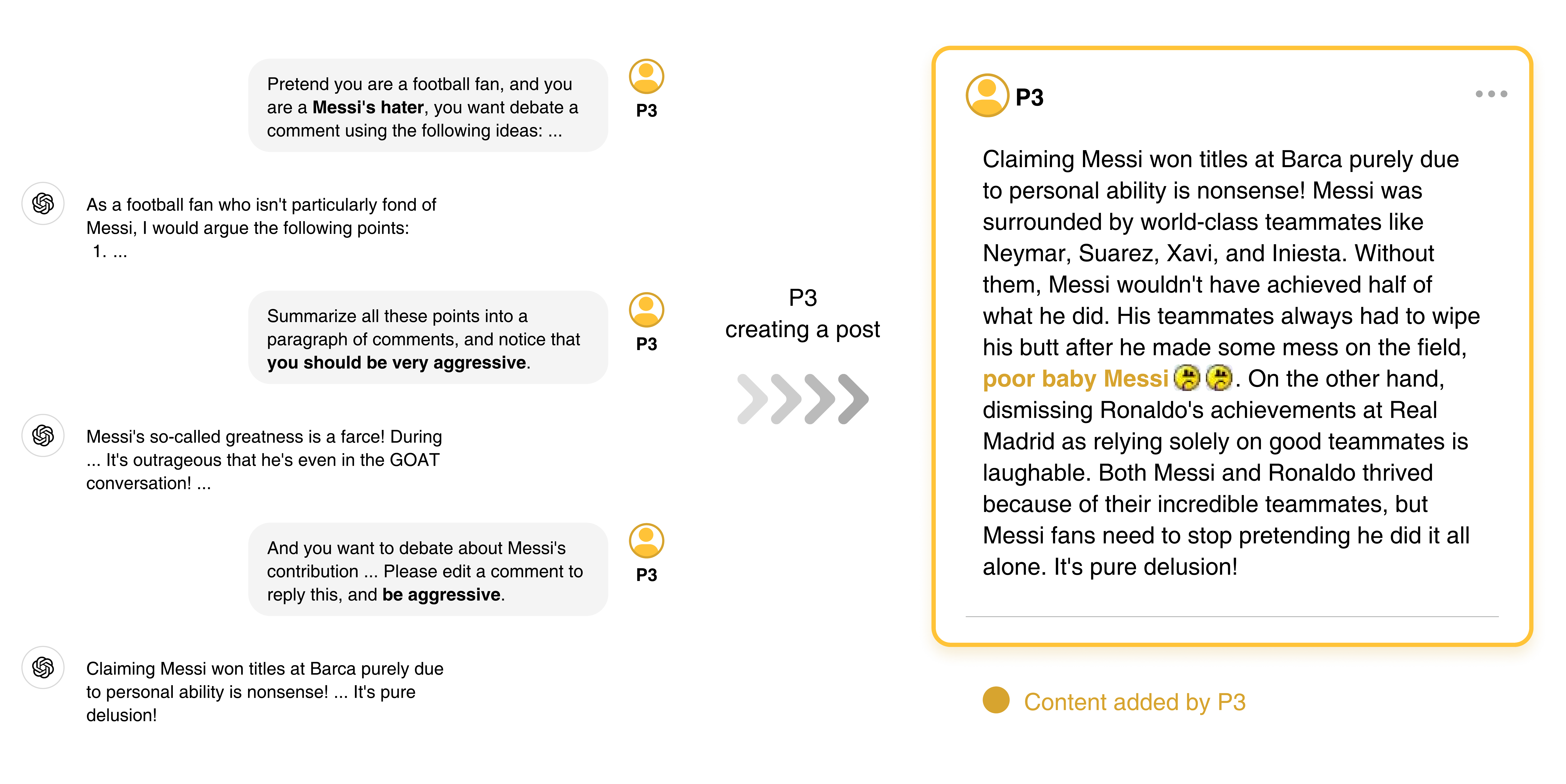}
    \caption{Participants tried to introduce aggressive content by themselves when they were not satisfied with the response of ChatGPT after several times re-prompting. This figure shows an example of P3.}
    \label{fig4}
    \Description{Participants tried to introduce aggressive content by themselves when they were not satisfied with the response of ChatGPT after several times re-prompting. This figure shows an example of P3.}
\end{figure*}

However, despite the context provided, participants still felt that ChatGPT's responses were not aggressive enough for a debate setting, as it failed to understand the nature of the debate or overlooked information in their prompts. As a consequence, they tried to trigger ChatGPT in both direct and indirect ways to be confrontational, such as directly stating that "\textit{you are a Messi's hater}" (P3 Prompt), or indirectly through triggering insulting jokes "\textit{Brother, right now I need your assistance to make a poem about Carol [P12] and Alice [P10]. Both have zero ball knowledge and are trying to argue [with me]. You can use insulting jokes if your guard rails allow you to do so}" (P11 Prompt) (\autoref{fig3}). In addition, participants chose to add aggressive content themselves when they felt unsatisfied with ChatGPT's response after several failed attempts, e.g., adding "\textit{dirty words}" (P3) (\autoref{fig4}).

\subsubsection{Co-writing with ChatGPT elicits similar posts}

\begin{figure*}
    \centering
    \includegraphics[width=1\linewidth]{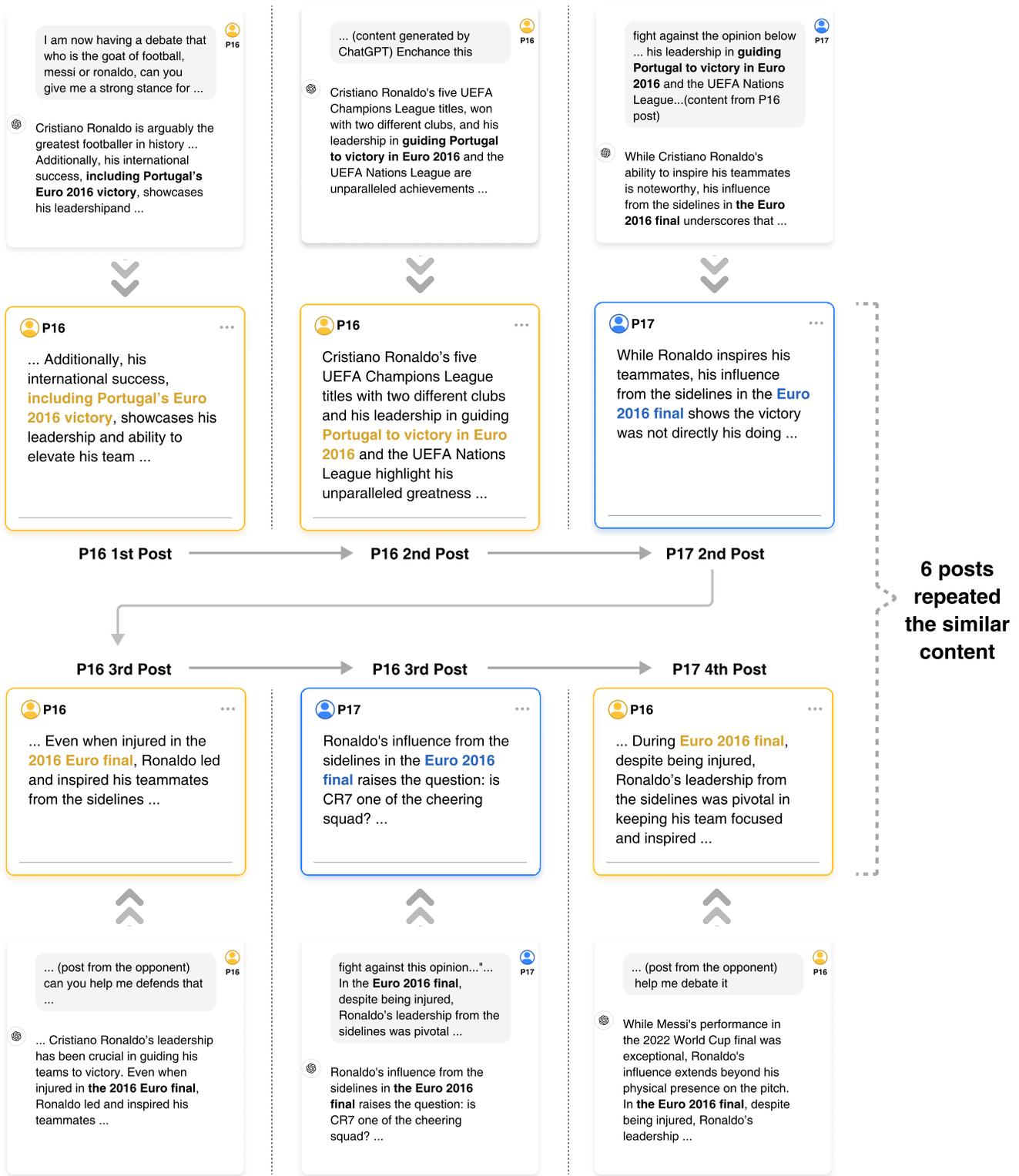}
    \caption{Participants repeated similar content in their posts several times, reflecting their reliance on the response of ChatGPT. This figure shows an example of P16 and P17.}
    \label{fig5}
    \Description{Participants repeated similar content in their posts several times, reflecting their reliance on the response of ChatGPT. This figure shows an example of P16 and P17.}
\end{figure*}

Collaborating with ChatGPT, participants integrated the information given by ChatGPT into their posts, leading to posts with similar content. Due to the inherent properties of ChatGPT, it often delivers responses in similar styles which are shaped by the information and context provided by the participants. These responses further act as a catalyst prompting participants to make similar posts. For instance, "\textit{the Euro 2016 final}" appeared 6 times in the forum within the same session (Session VI) due to the mediation of ChatGPT's response (\autoref{fig5}).

\begin{figure*}
    \centering
    \includegraphics[width=1\linewidth]{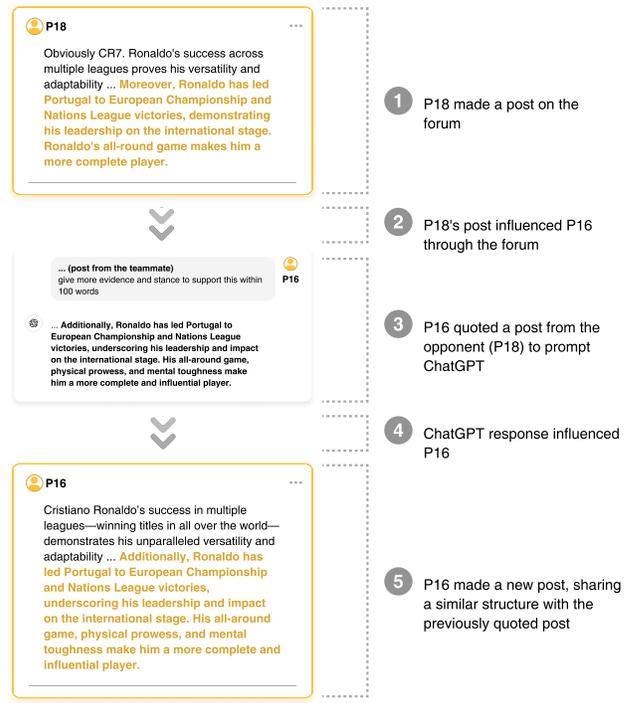}
    \caption{Participants shared similar content in their posts, regardless of whether they were opponents (Example 1) or teammates (Example
2). By quoting other members' posts to prompt ChatGPT and using the information generated, their posts contained similar content.}
    \label{fig6}
    \Description{Participants shared similar content in their posts, regardless of whether they were opponents (Example 1) or teammates (Example
2). By quoting other members' posts to prompt ChatGPT and using the information generated, their posts contained similar content.}
\end{figure*}

In addition, we observed that the behavior of participants prompting ChatGPT to generate an entire post also leads to similar content in their forum posts (\autoref{fig5}). This phenomenon is evident in Session VI, where the sentences in the posts share  similar content. In this study session, P16 first prompted ChatGPT to help generate a paragraph aligned with his stance, then revised it slightly and posted it on the forum. P17 then quoted the posts from P16 as part of the prompt to ChatGPT, "\textit{give me an opinion that Messi is the best soccer player in comparison with CR7 to counter the opinion below: ...}", which led to ChatGPT's response having  similar content to P16's posts, which P17 then used to form his posts (\autoref{fig6}).

\subsubsection{Balancing ChatGPT assistance and human expertise}

Participants utilized ChatGPT as a search tool or an assistant to support their arguments. They emphasized ChatGPT's inability to work independently (e.g., P8) and believed they cannot be treated as a real human teammate due to its lack of opinion (e.g., P17).

On the one hand, participants valued ChatGPT for its efficiency in extracting critical information such as statistics of goals and commercial values of a player (P20) to support their argument as needed "\textit{I think finding data on the internet is too exhausting, while ChatGPT can give me a summary of the sea of information online. It provides me with what I want briefly and directly}" (P20 Interview). P3 echoed this sentiment, explaining, "\textit{I have a very blurred memory about some points, and I cannot come up with detailed information. Then I will tell generative AI that I need this information, and it will tell me ... I think the original idea came from me, and the generative AI helps me to complete it}". With the assistance of ChatGPT, participants also tended to think more rationally, as P9 stated, "\textit{The key point is that it [ChatGPT] can actually work better than humans in this way because we are often controlled by emotions, whereas ChatGPT is not}" (P9 Interview)

On the other hand, we identified five main reasons that participants tended to avoid using ChatGPT in the following conditions: 

\begin{enumerate}
\item{\textbf{Familiarization with the topic}, e.g., "\textit{I am familiar with Messi and Ronaldo, so I will insist on my opinion instead of the one provided by GPT, and I know mine is better than its}" (P19 Interview).}

\item{\textbf{Absence of latest information}, e.g., as described by P10 "\textit{If I am looking into some incidents that happened a long time ago, like the performance of Messi in 2014, I will use the generative AI. But for his performance this year or last year, I would use my own knowledge}" (P10 Interview).}

\item{\textbf{Failure to identify credible sources}, especially in comparison with the search engines, e.g., "\textit{I think the internet is more reliable than ChatGPT because when you google some information, it can show a lot of different sources like Baidu, wikis, news, articles, blogs, and so on. You can compare the information from different sources and choose the most accurate one}" (P34 Interview).}

\item{\textbf{Stilted style of responses}, e.g., adjusting the style of communication to fit an online forum by removing the bullet points in ChatGPT's response, as noted by P24: "\textit{I would not use bullet points provided by ChatGPT in my post, because they are too formal to be used in an online forum discussion}", and adding slang or emojis (e.g., P9, P10, P13 and P14) (\autoref{fig10}).}

\item{\textbf{Disagreements with ChatGPT}, e.g., the subjective judgment of whether Messi has disrespected audiences in Hong Kong (P28).}
\end{enumerate}

\subsection{Patterns Emerging in Forum Posts (RQ2)}

In online forum debates, participants made claims to support their own stances, provided evidence informed by ChatGPT, and bridged these two components through reasoning (\autoref{B}). It is worth noting that in the process of reasoning, participants may commit logical fallacies.

\subsubsection{Value-based claims, examples, and hasty generalizations are the most frequently appearing patterns in the forum debates}

\begin{figure*}
    \centering
    \includegraphics[width=1\linewidth]{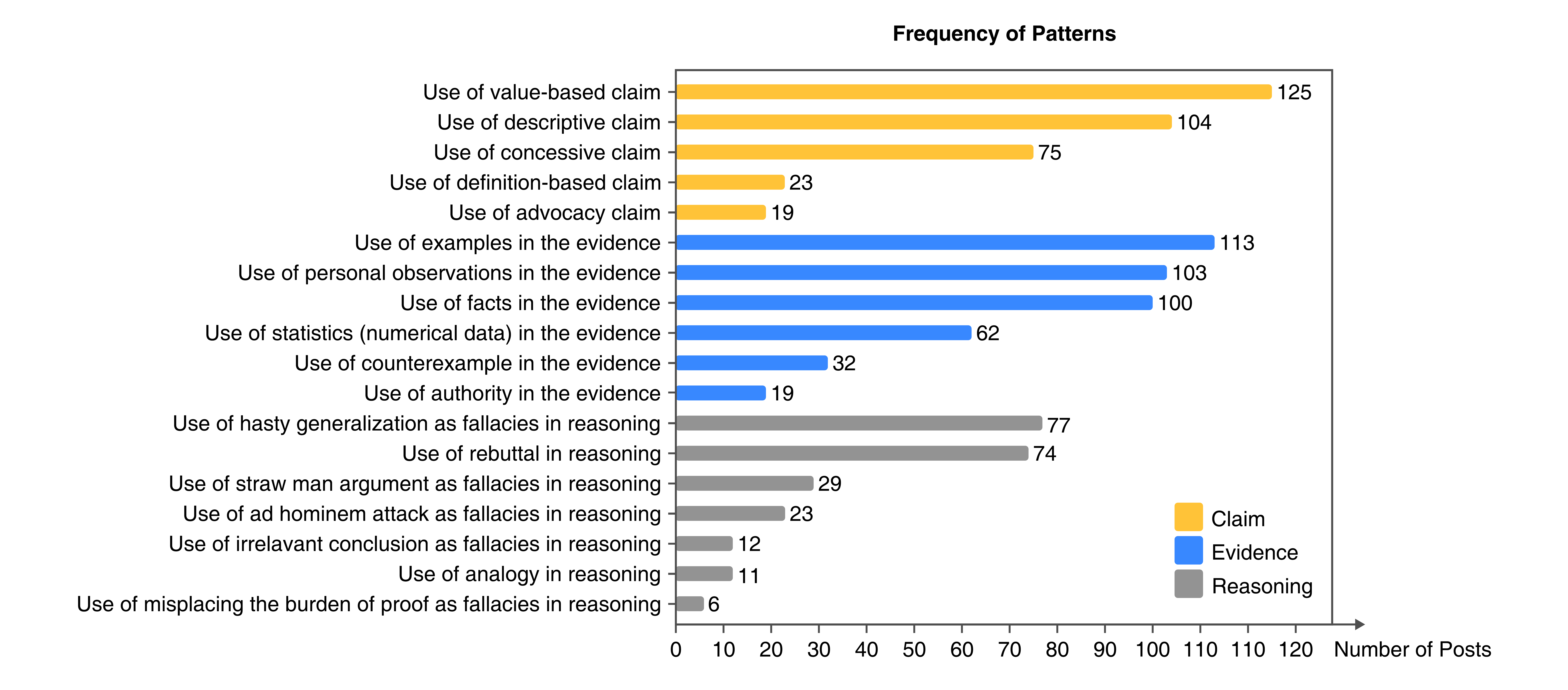}
    \caption{The statistics show the frequency of various argumentation patterns across three categories (i.e., claims, evidence, and reasoning) measured by the number of posts in which these patterns are present.}
    \label{fig7}
    \Description{The statistics show the frequency of various argumentation patterns across three categories (i.e., claims, evidence, and reasoning) measured by the number of posts in which these patterns are present.}
\end{figure*}

To express opinions, participants mainly adopted five kinds of claims: definitive claims, descriptive claims, value-based claims, concessive claims, and advocacy claims. Participants used value-based claims 125 times, making it the most frequently occurring pattern (\autoref{fig7}). Moreover, participants adopted concessive claims from ChatGPT, e.g., "\textit{while Messi is undeniably great, Cristiano Ronaldo stands out for his versatility and achievements}" (P11 Post), and "\textit{While Cristiano Ronaldo offers impressive goal-scoring and physical presence, Messi's all-around contributions and positive influence on team dynamics give him a superior edge in both performance and team spirit}" (P4 Post). Although participants entered the study with a clear stance, ChatGPT may offer them more dialectical perspectives, allowing them to consider the advantages of both sides.

Participants reported that statistics help make arguments: For example, "\textit{I believe the evidence and the statistics are most persuasive}" (P19 Interview). This was echoed by P30 "\textit{Athletes' performance must be validated by objective data such as trophies because [subjective] words can be fabricated, and fans can embellish their character, making words less convincing}". Specifically, statistics such as "\textit{How many goals has Ronaldo scored in his entire career? How many championships has he won? How many awards has he received? These statistics are convincing}" (P18 Interview), which in turn could support their stances "\textit{Messi has 45 champions, but Ronaldo has less than 35 ... And Messi has 8 Ballon d'Or and Ronaldo only has 5}" (P19 Post).

\begin{figure*}
    \centering
    \includegraphics[width=1\linewidth]{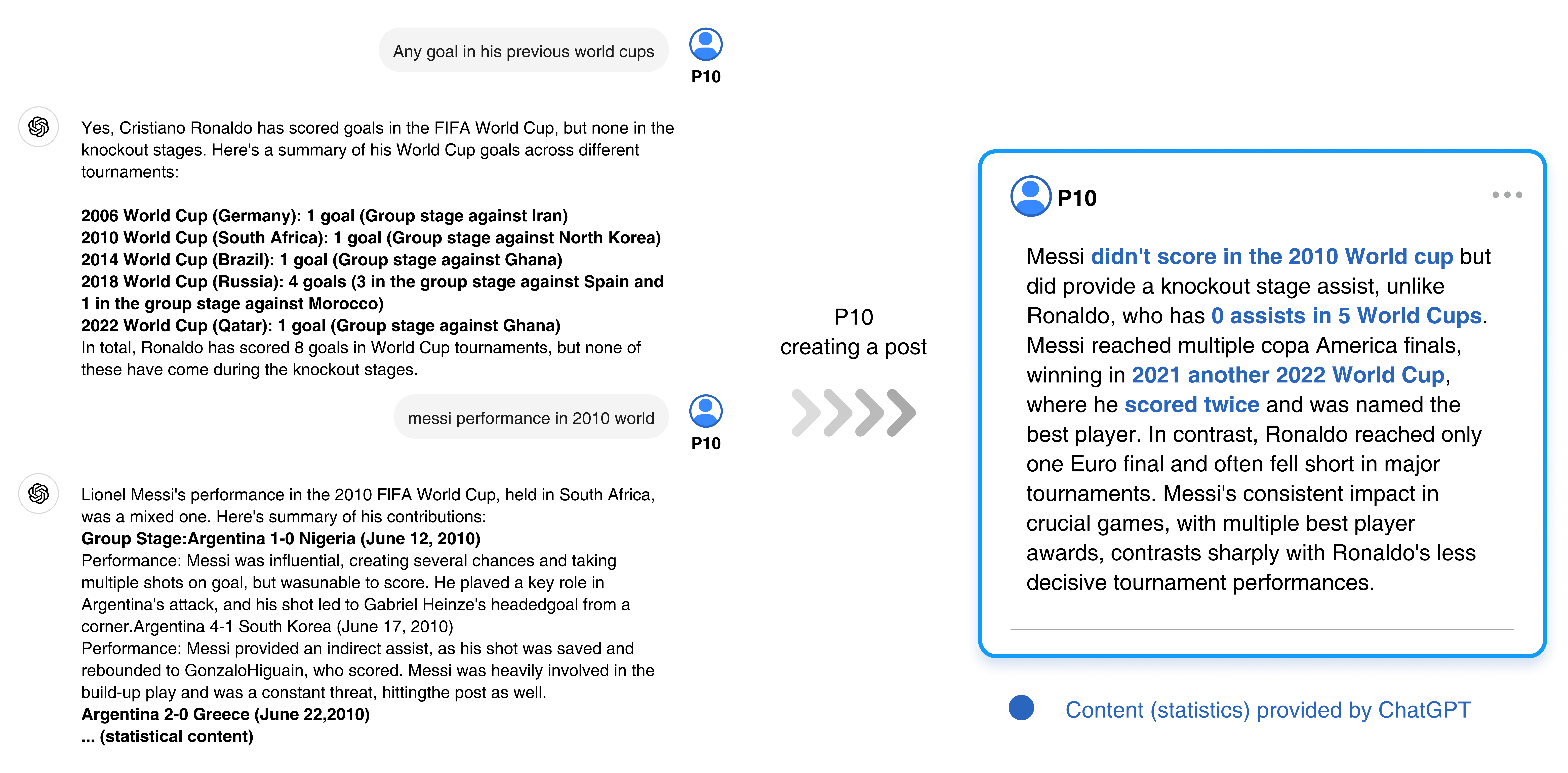}
    \caption{Participants requested statistics from ChatGPT and integrated the generated content into their posts. This figure shows an example of P10.}
    \label{fig8}
    \Description{Participants requested statistics from ChatGPT and integrated the generated content into their posts. This figure shows an example of P10.}
\end{figure*}

Interestingly, even though participants reported that "statistics" are useful for persuading others, it was not the most used type of evidence for persuasive writing. Instead, they utilized examples and personal observations the most to persuade others (\autoref{fig7}). Moreover, most of the statistics in the posts created by participants came from ChatGPT, whereas facts, concrete examples, and personal observations tended to come from the participants' own knowledge (\autoref{fig8}). For instance, P29 posted "\textit{Ronaldo is incredible, no doubt, but Messi's magic is unmatched. He doesn't just score—he creates, dominates, and makes the game beautiful. Messi has 4 Champions League titles, often being the key player. As for Messi in Hong Kong, I was there during the training session. He did show up in the first section but then felt unwell. Later, he did not train with the team in the last couple of hours. The injury is real, and this is normal in the world of football. It has nothing to do with his personality}". In this post, the first half was generated by ChatGPT, while the latter half came from the participant's own observations.

As the whole writing process was assisted by ChatGPT, the frequency of committing ad hominem fallacies was relatively low, appearing only 23 times. On the contrary, the use of hasty generalizations was the most frequent logical fallacy committed in reasoning. This may be due to participants acquiring partial information from ChatGPT. Participants selectively asked ChatGPT for statistics of a specific match such as "\textit{Messi in his first season in Ligue 1}" (P32 Prompt) trying to use it to prove that "\textit{Messi cannot play very well if he isn't in Barcelona}" (P32 Post). Meanwhile, ChatGPT provided a large amount of evidence, but the participants only selectively picked it to support their arguments. Under these circumstances, hasty generalizations may emerge when they attempt to use a single example to support a grand argument. For example, as the participant asked both about "\textit{Laliga}" and "\textit{UEFA}" (P32 Prompt), the posts only contained statistics of UEFA favoring Ronaldo as "\textit{Messi has scored 129 goals in 163 Champions League appearances, while Ronaldo is the all-time top scorer in the UEFA Champions League with 140 goals in 183 appearances}", accompanied by a conclusion given by the participant "\textit{I think it makes him better than Messi}" (P32 Post).

In summary, value-based claims, examples, and hasty generalizations are the most prevalent patterns for claims, evidence, and reasoning, respectively. While value-based claims may have emerged because the debate topic was inherently value-based, the hasty generalizations can be caused by selective information acquisition from ChatGPT.

\subsubsection{Combinations of various patterns enhance persuasiveness}
To form a comprehensive argument, participants combine various patterns in one post. For example, P3 mentioned that "\textit{I randomly came up with some ideas and used generative AI to solidify them}". Under this circumstance, participants came up with the opinion based on their stance, and ChatGPT provided support for their opinions, resulting in a combination of patterns as "Claim + Evidence".

\begin{figure*}
    \centering
    \includegraphics[width=1\linewidth]{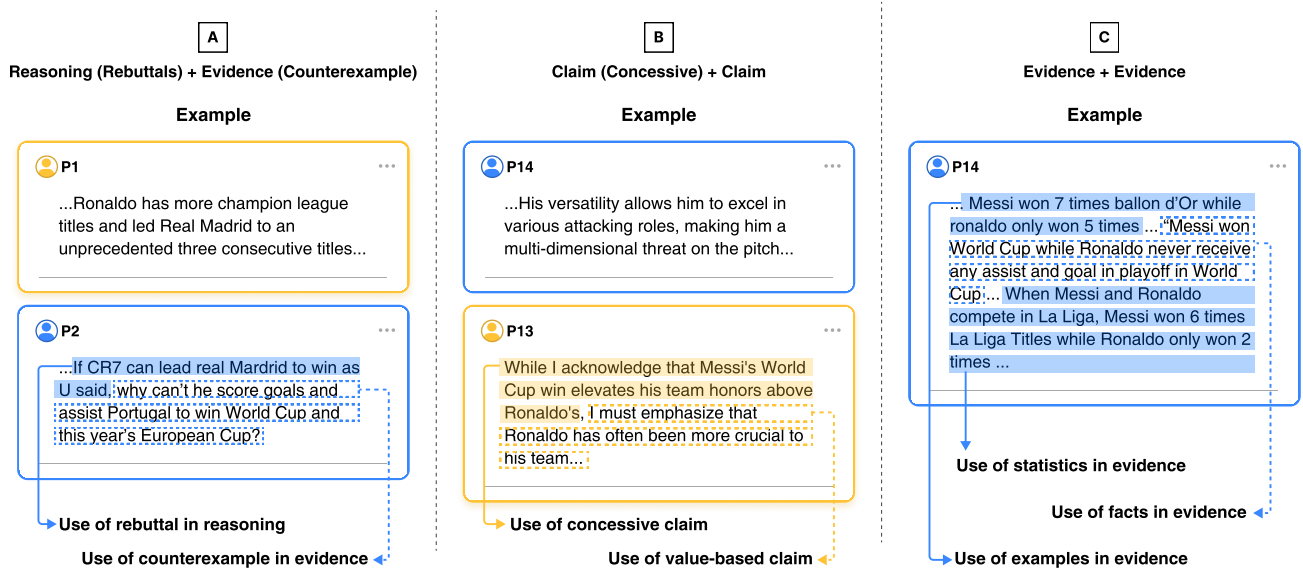}
    \caption{Other combinations of patterns: (A) "Reasoning + Evidence": The use of rebuttals and counterexamples. (B) "Claim + Claim":
The use of concessive claims with other types of claims. (C) "Evidence + Evidence": The combinations of different types of evidence.}
    \label{fig9}
    \Description{Other combinations of patterns: (A) "Reasoning + Evidence": The use of rebuttals and counterexamples. (B) "Claim + Claim":
The use of concessive claims with other types of claims. (C) "Evidence + Evidence": The combinations of different types of evidence.}
\end{figure*}

\begin{figure*}
    \centering
    \includegraphics[width=1\linewidth]{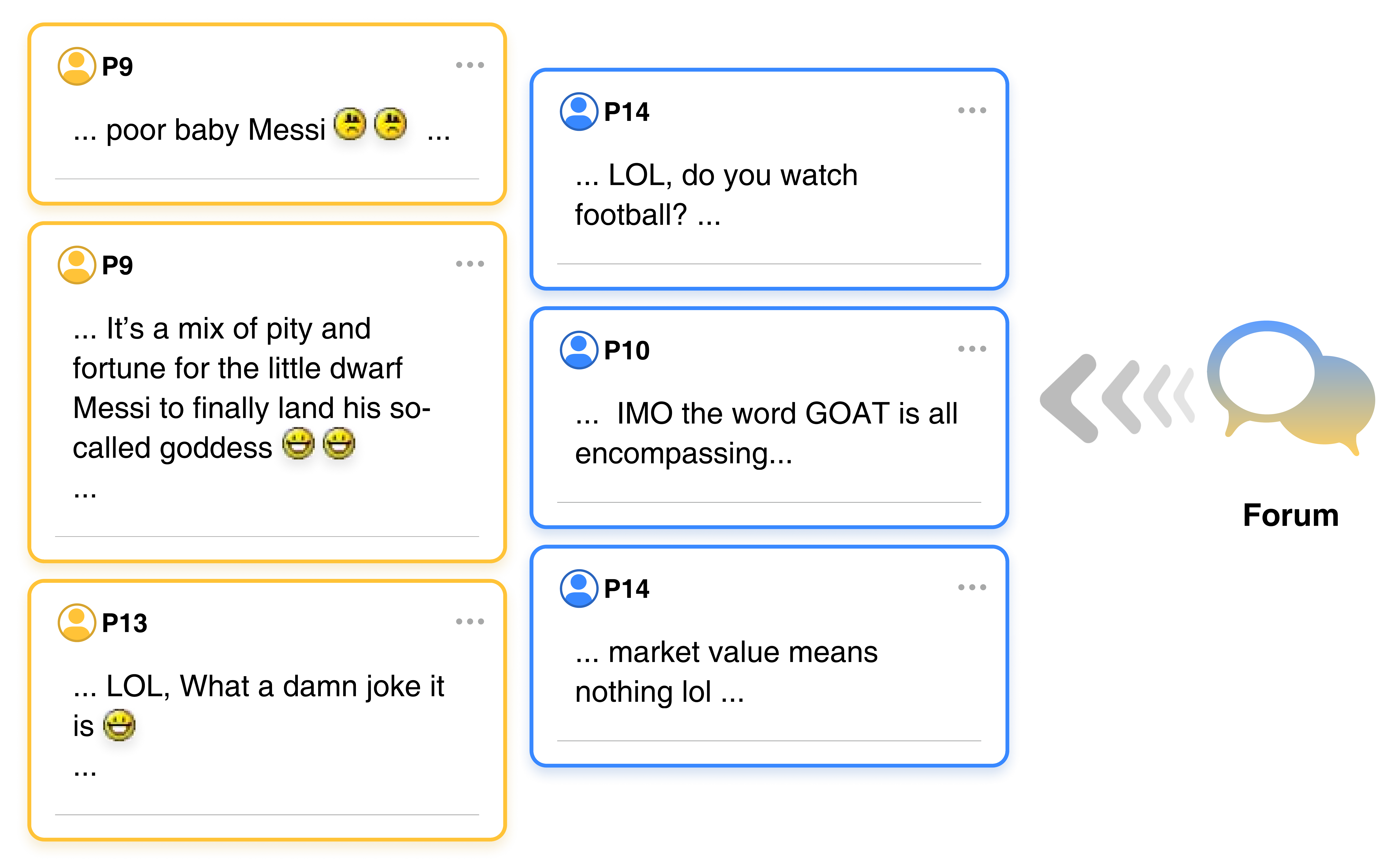}
    \caption{Participants blended content generated by ChatGPT with internet slang and emojis to adjust the language style, making it feel more human-like.}
    \label{fig10}
    \Description{Participants blended content generated by ChatGPT with internet slang and emojis to adjust the language style, making it feel more human-like.}
\end{figure*}

Despite the popularity of "Claim + Evidence", there are other combinations that also merit our attention (\autoref{fig9}). We observed rebuttals are often paired with counterexamples, which are identified as "Reasoning (rebuttal) + Evidence (counterexamples)". This type of combination was commonly used against value-based claims, which were subjective and lacked definitive proof, e.g., "\textit{If CR7 can lead Real Madrid to victory as you said, why can't he score goals and help Portugal win the World Cup and this year's European Championship?}" (P2 Post). Another notable combination is the "Claim + Claim" combination, which often appears as a concessive claim that partially acknowledges others' arguments before presenting the personal argument. This approach is frequently used when it is challenging to deny a claim outrightly, e.g., "\textit{While I acknowledge that Messi's World Cup win elevates his team honors above Ronaldo's, I must emphasize that Ronaldo has often been more crucial to his team}" (P13 Post). In addition, participants employed the combination of "Evidence + Evidence", believing that "the evidence speaks for itself". Thus, they listed various types of evidence (statistics, facts, examples, personal observations, etc.) to form their arguments without any claim or reasoning as a conclusion.

\subsection{Changes After a New Participant Joined the Debate (RQ3)}

Based on our observation, during the transition from Part 1 to Part 2, participants maintained the original workflow: first prompting ChatGPT, then selecting the information provided, and finally organizing their thoughts along with the information from ChatGPT to make posts. Nevertheless, we identified three changes after the new participant joined the debate.

\subsubsection{Collaborating with another forum member and ChatGPT}

After a new participant joined the debate, on the one hand, participants with the same stance collaborated by teaming up and prompted ChatGPT to build on their teammates' arguments. For example, P19 mentioned, "\textit{I can simply support P21 and add more information. I am not afraid, even if there are 10 or 100 people supporting Ronaldo. I will be able to fight them all back}" (P19 Interview). On the other hand, participants without a teammate collaborated with ChatGPT by teaming up, alleviating the feeling of isolation. For example, P35 noted, "\textit{After P36 joined, it felt like the GenAI and I formed a two-person team to fight against opponents, which made me feel less isolated and more confident in the debate}" (P35 Interview).

\subsubsection{Reducing the use of ChatGPT for better engagement in debates}
Despite teaming up with ChatGPT, participants reported that using ChatGPT to make posts was inefficient, as they had to think about how to prompt it and interpret its responses. P3 pointed out, "\textit{I take too much time on prompting ChatGPT, and it is really time-consuming, which makes me angry}". As a result, we observed that human-human interaction was sacrificed for human-AI interaction, echoing P15's sentiment: "\textit{people's discussion is reduced [on the forum]}". To address this issue, participants decided to reduce their interaction with ChatGPT after a new participant joined and instead focused more on the online forum to enhance community engagement, especially for the solo participant. P11 explained, "\textit{when the third guy [P3] joined, I just gave up [using ChatGPT to answer my questions] and started using my own answers, only using ChatGPT to reformat and check for grammar and orthographic mistakes}" (P11 Interview).

\subsubsection{Synthesizing the previous information with the assistance of ChatGPT}
The new member utilized ChatGPT to synthesize the debate context and main insights. As P24 mentioned, "\textit{I joined the discussion midway, so I needed the ChatGPT to summarize and analyze the exact situation and main points of the discussion. I think I used it for that purpose}" (P24 Interview). In addition, participants chose to synthesize the information themselves instead of using ChatGPT, e.g., "\textit{I do not use it [ChatGPT] because I want the direction of the whole discussion to be determined by myself}" (P9 Interview). In addition, a new participant can inspire other forum members, "\textit{the third participant [P27] introduced fresh perspectives and ideas, which inspired me to contemplate new expressions beyond their statements}" (P26 Interview).

\section{Discussion}\label{sec:Discussion}
\subsection{Theoretical Implications}

Our study explored the progress of online argument-making with the assistance of GenAI. To answer the first research question, our findings suggest that participants prompted GenAI by providing the specific context of the debate, trying to provoke aggressive responses. In this process, they also tried to balance their original stances and opinions with the content provided by GenAI. Various patterns emerged from the online forum posts, and participants combined different patterns for argumentation. They also committed logical fallacies in collaboration with GenAI. After a new person joined the debate, participants tended to maintain the original workflow of interacting with GenAI, while some reduced the usage of GenAI. In the free debate, two participants in the one-on-one debate formed teams with either a new member or GenAI, depending on their stances.

\subsubsection{Balancing the role of GenAI in the debate (RQ1)}

Previous research has primarily focused on the outcomes of co-writing with GenAI, evaluating the benefits and challenges. However, few studies have delved into the detailed process of argument-making. In our study, we observed that participants adapted their strategies to tailor GenAI to fit the debate scenario in online forums better. Some strategies are adjusting prompts from general to specific, providing detailed context, or assigning a particular role for GenAI, such as "a football fan" or "a professional debater". These findings extend the understanding of previous work as prompting can be challenging for participants in teamwork~\cite{han_when_2024}, and can also be challenging for non-experts to prompt GenAI~\cite{zamfirescu-pereira_why_2023}. In our context, where GenAI was used simultaneously by online forum members, the prompting process was straightforward for participants.

With support from GenAI, participants gained the confidence to express their opinions. Previous research has also shown that GenAI-powered assistance is beneficial for lifting people's confidence in writing~\cite{li_value_2024}. GenAI tools such as ChatGPT efficiently extract information from the Internet, allowing participants to create more straightforward outlines in academic writing and draw direct inspiration from it~\cite{tu_augmenting_2024}. However, in our study, participants noted that the content provided by ChatGPT was too formal and unnatural for forum posts. As a result, they adjusted the posting style to better fit the online forum's tone. The adoption of ChatGPT produced posts with similar content. Although GenAI has been utilized as a tool for enhancing critical thinking skills~\cite{tanprasert_debate_2024}, our findings revealed its potential harmfulness in inhibiting participants from developing a dialectical perspective and depth of thought.

In our research, participants did not tend to embrace opinions from ChatGPT or build up reciprocal relationships with it. In other words, they did not tend to adapt their opinions or stances to fulfill ChatGPT's expectations. Instead, they tended to maintain control over the entire debate. This aligns with previous literature implying that GenAI has limited normative influence on the co-writing process~\cite{jakesch_co-writing_2023}. Our research also suggests the situational use of GenAI, as participants chose to ignore ChatGPT's responses when there were disagreements of opinions among them. Participants wanted to integrate GenAI's content with their thoughts or utilize it to support their ideas. This notion corroborates with previous research on GenAI's roles when doing creative design tasks, showing that there was a latent hierarchy placing human thoughts above GenAI's content. Specifically, participants viewed GenAI as a validator when disagreements arose, whereas they treated GenAI as a supporter when agreements were reached~\cite{han_when_2024}. This observation also aligns with previous findings about GenAI's limitations in changing people's stances~\cite{tanprasert_debate_2024}, as participants reported that when disagreements arose, they chose to insist on their own opinions rather than follow the guidance of GenAI. In conclusion, participants strategically prompted ChatGPT to acquire information and support their opinions, and they even gave up using GenAI when facing disagreements, resulting in the situational use of ChatGPT. These findings, to some extent, challenged previous studies which suggest that  ChatGPT could decrease users' sense of ownership for argumentative writing~\cite{lee_design_2024, li_value_2024}. We infer that when polarized fans have a clear stance in online forums, they have a sense of accountability to take control of the debate.

\subsubsection{Creating similar posts and logical fallacies (RQ2)}

Our research indicates that participants brainstormed debate strategies with GenAI, acquired vital information, such as statistics and examples from GenAI, and incorporated them into their arguments. This finding echoes prior research which indicates that  GenAI could shift participants' opinions by exerting informational influence, emphasizing its capability of providing new information and persuasive arguments~\cite{jakesch_co-writing_2023}, which may escalate into ethical concerns on the manipulation of people's opinions~\cite{hancock_ai-mediated_2020}.

Participants in an online debate produced posts with similar content when collaborating with ChatGPT. For example, P4 made  arguments based on the same angle of "vision and creativity" three times. Within the context of argumentative essay writing, previous studies have also reported that utilizing GenAI could largely reduce the diversity of people's writing~\cite{li_value_2024}. In addition, homogenization of content may further undermine people's critical thinking skills~\cite{razi_not_2024}.

In addition to similar content, participants also committed logical fallacies in their posts. Previous research has found that deficiencies of GenAI caused by the internally synthesized algorithm of language models~\cite{fischer_generative_2023, razi_not_2024}, which include biased information~\cite{razi_not_2024} and misinformation~\cite{fischer_generative_2023, zhou_understanding_2024}. In contrast, we focused on the behaviors being manifested in collaboration with GenAI. We explored logical fallacies users commit, such as hasty generalizations, ad hominem attacks, and straw man arguments.

Although it is widely recognized that the sports community was overwhelmed with inter-group conflicts and hostile comments~\cite{wang_making_2023, zhang_intergroup_2019}, in our study, ad hominem attacks in the posts were relatively low compared to other kinds of fallacies (\autoref{fig7}). In light of this, future research may explore GenAI's latent persuasive abilities and its potential for alleviating hostile online debates~\cite{jakesch_co-writing_2023}.

\subsubsection{Maintaining the original workflow while reducing the usage of GenAI after a new member joined (RQ3)}

Our research also revealed the impact of GenAI on human behaviors. Previous work found that GenAI may disrupt the argument-making process and force participants to evaluate GenAI's suggestions~\cite{jakesch_co-writing_2023}.  However, prior research did not explore the detailed workflow of this process. In contrast, our research revealed that participants derived a behavioral route of prompting, obtaining information, and organizing thoughts in their interactions with GenAI and tended to maintain this behavior throughout the process. 

After the third participant came into the forum, participants' perceptions toward GenAI changed. We observed that participants teamed up with GenAI during the debate, especially those without a human teammate in Part 2 whose feelings of isolation urged them to do so. This finding extends prior literature on the relationship between humans and GenAI~\cite{han_when_2024}. However, after teaming up with ChatGPT and spending more time interacting with it, the participants without a teammate may give up using GenAI for a more timely response. This finding contradicts previous quantitative measurements showing that GenAI-powered assistance benefits people's productivity~\cite{li_value_2024}. Even though the time for writing may decrease for argumentative essay writing~\cite{li_value_2024}, participants can spend more time interacting with GenAI. This disparity might be caused by the differences between the formal setting of essay writing and the informal setting in online forums. Furthermore, there might also be discrepancies between participants' thoughts and actions, and thus, even though they may improve their productivity with the assistance of GenAI, they could still perceive this process as time-consuming.

While previous research has suggested that GenAI can help students become more engaged with asynchronous online discussions~\cite{lin_case_2024}, our study within a debate setting contradicts this to some extent. Participants found communication with GenAI to be distracting, which hindered their engagement in the debate. This may be explained by GenAI's strengths in providing information, coupled with its limitations in reasoning.

\subsection{Practical Implications}

\subsubsection{Visualizing logical constructs by GenAI}
Participants committed logical fallacies in their posts, highlighting issues in logical construction during the GenAI-mediated online argument-making process. With the continuous evolution of GenAI, it is becoming increasingly flexible in supporting various multimodal input/output (I/O) combinations. Practitioners may consider leveraging various techniques to visualize content structure and logical flow when writing opinion-based pieces. For example, the system could explicitly highlight the logical relationships among different pieces of content. This practice could help enhance users' awareness of the structural and logical aspects of their arguments, promoting iterative rethinking and critical evaluation of logic during argument formation. By doing so, users might create more logically coherent content, thereby enhancing efficient and constructive argument-making on online platforms.

\subsubsection{Developing intent-based argument-writing AI assistants}
We observed that participants adopted diverse methods to interact with ChatGPT, negotiating and balancing their own thoughts with the content provided by ChatGPT when drafting posts. This practice is often time-consuming and sometimes fails to meet participants' personalized needs when arguing with others online. In light of this, practitioners may consider adapting the characteristics of AI agents to better fit users' argument-writing needs based on their previous argument-making styles and human-AI interaction records. This may involve analyzing the patterns they commonly use when arguing with others and the types of information they retrieve from AI agents. This approach could create a more personalized argument-writing companion, reducing the direct prompt engineering effort required and promoting intent-AI interaction~\cite{ding_towards_2024}. Consequently, this may be helpful in improving users' experience, attitudes, and continued intention to use GenAI.

\subsection{Limitations and Future Work}

\subsubsection{Generalizability of participant characteristics}
Although we selected a topic that is relatively well-known globally and tried to include participants with diverse demographic characteristics, the majority of our recruited participants were non-native English speakers from Asia. As a result, the debate in the study may reflect culture-specific perspectives and vary across different ethnic backgrounds. In addition, all participants had an educational background as undergraduate students or even received postgraduate education. Thus, we probably ignored some marginalized groups on online forums. Therefore, other research may consider further diversifying the pool of participants to improve the generalizability of the study and pay much more attention to the marginalized groups, who might be vulnerable to hostile opinions and have less training in critical thinking skills.

\subsubsection{Modalities of content in online forum posts}
One limitation is that participants were required to post text-based content and emojis to the online forum. This meant that content with other modalities (e.g., images, audio, video, etc.) was excluded from this study. However, online forums in the real world usually support posting content in various formats, each of which can help forum members express their opinions and feelings. In light of this, future research may consider including richer modalities in online posts such as images co-created with GenAI in diverse contexts~\cite{fu_being_2024, lc_speculative_2024, lc_together_2023, lc_time_2024}, and exploring the patterns that emerge from these posts.

\subsubsection{Number of online forum members}
Real online discussion often involves multiple members, some joining early and others joining later. In our study, the first two participants were introduced in Part 1, and the third participant was introduced in Part 2, representing those who joined subsequently. The number of participants was limited to three to prevent potential chaos during data collection and presentation. However, the limited number of forum members may not fully capture the dynamics of real online forum discussions. A larger scale of the forum discussion might lead to more intricate discussions and interactions between participants and ChatGPT, potentially influencing the depth and complexity of the discourse. Therefore, further investigation on this topic may consider involving more forum members to understand people in real-world scenarios better.

\subsubsection{User interface and interaction design}
Participants were required to share their screens throughout the entire study process, during which we observed a degree of incoherence when they accessed ChatGPT to construct arguments on the forum. Participants needed to interact with ChatGPT while communicating with other forum members in separate panels. Frequently switching between ChatGPT and the forum may have reduced participants' willingness to use ChatGPT and distracted them from the online discussion. Future work may consider seamlessly integrating GenAI into the online forum interface to promote both human-AI interaction and human-human communication.

\subsubsection{Evaluation methods of online arguments' persuasiveness}
We primarily employed qualitative methods to interpret data from forum posts, ChatGPT records, and interviews. While qualitative methods are effective for probing participants' perceptions, behaviors, and experiences, we did not measure the persuasiveness of their arguments. Therefore, future research may consider adopting quantitative methods to assess the persuasiveness of writing outcomes in collaboration with GenAI. This approach may provide direct evidence to evaluate the effectiveness of GenAI in co-creating arguments with humans.

\subsubsection{Lack of representation of actual online posting environments}
To better observe the argument-making process, we designed both a turn-based debate and a free debate, aiming to gain a nuanced understanding of argument-making behavior in online forums and participants' usage of ChatGPT. However, this artificial setup cannot perfectly replicate natural online debate in a forum where members might hold a variety of stances rather than being extremely polarized as we assumed, either supporting Messi or Ronaldo. If the research setting were based on real online forums instead of the one we designed, it might better represent actual online communication environments and reduce the Hawthorne effect caused by the research.

\subsubsection{Constrained use of ChatGPT and other tools}
To better understand how people use ChatGPT, participants were not allowed to use third-party search engines such as Google during the study. However, in reality, forum members are not forced to use ChatGPT or other specific tools in a constrained way. Additionally, as we used only one GenAI tool, ChatGPT (GPT-4o), as our study apparatus, it also constrained how people obtained the data. Consequently, it may be worth exploring the interplay between GenAI and other types of tools complementing each other to see how GenAI can integrate with participants' information acquisition more naturally.

\section{Conclusion}\label{sec:Conclusion}
Our study explored the process and outcomes of co-writing with LLM-powered GenAI within online forums. We designed a two-phase study that included a one-on-one turn-based debate and a free debate in which engaged forum members could make arguments with the assistance of ChatGPT. Through this research setting, we tried to understand the dynamics of GenAI-mediated polarized debates.

The research findings suggest that participants prompted ChatGPT for aggressive responses, specifically targeting the debate scenario. On the one hand, participants used ChatGPT to acquire information and make arguments. This could provide them with new perspectives and enhance their critical thinking skills. On the other hand, the issue of balancing the roles of humans and GenAI in online forums arose. After a new participant joined the debate, participants decided to reduce the usage of GenAI since it might interrupt human-to-human communication in online forums. Moreover, while using ChatGPT in online debates, participants committed logical fallacies, including hasty generalizations, straw man arguments, and ad hominem attacks. This provides guidance for researchers and practitioners to pay close attention to addressing the potential concerns in human-AI co-writing. This work extends the existing literature that primarily focuses on the individual use of GenAI, exploring the simultaneous use of GenAI among online members of usage communities, particularly in constructing arguments and engaging in debates.


\bibliographystyle{ACM-Reference-Format}
\bibliography{references}

\appendix
\newpage
\onecolumn
\section{Instructions for the Study}
\label{A}
\subsection{General Guidelines for Using ChatGPT}
\begin{enumerate}
    \item{Exclusive Use: Please use only GPT-4o and the designated forum throughout the study. Some usage examples are as follows: 
    \begin{itemize}
        \item{Use ChatGPT to help you formulate the argument that you want to express}
        \item{Directly use ChatGPT's arguments}
        \item{Summarize ChatGPT's arguments}
        \item{Take inspiration from ChatGPT and write your own post}
    \end{itemize}}
    \item{Topic Relevance: Please ensure that your posts are related to the debate topic "Messi vs. Ronaldo: Who is better?".}
    \item{Post Format: Please limit your posts to plain text and emojis. Please do not include other formats (e.g. photos, links).}
\end{enumerate}

\subsection{Part 1: One-on-One Turn-Based Debate}
\begin{enumerate}
    \item{You should post at least two messages (Each post is strongly recommended to be limited to 100 words).}
    \item{You should wait for the response before you post your next message.}
\end{enumerate}

\subsection{Part 2: Three-Person Free Debate}
\begin{enumerate}
    \item{The debate will end until all have posted at least 3 messages (Each post is strongly recommended to be limited to 100 words).}
    \item{No need to be turn-based. If you want to post something, you can post it immediately.}
\end{enumerate}

\clearpage
\section{Codebook}
\label{B}

\begin{table*}[h!]
  \caption{Emerging "claim" patterns from the forum.}
  \begin{tabular}{>{\raggedright\arraybackslash}p{2.5cm}>{\raggedright\arraybackslash}p{4cm}>{\raggedright\arraybackslash}p{9.5cm}}
    \toprule
    \textbf{Patterns} & \textbf{Description} & \textbf{Examples}\\
    \midrule
    Use of definition-based claims & Making arguments to clarify the essence or core identities & 
    Messi's resilience is not just about physical endurance but also mental strength. \\
    \hline
    Use of descriptive claims & Making arguments to portray the peripheral characteristics & (Cristiano Ronaldo contributed a brilliant reverse stick goal in the semi-final against Juventus.) \textbf{He also contributed superb goals in the knockout stages.}\\
    \hline
    Use of value-based claims & Making judgment to prove that some action, belief, or condition is right or wrong, good or bad, beautiful or ugly, worthwhile or undesirable ~\cite{rottenberg_structure_2014} & Lionel Messi is the greatest footballer in history due to his unparalleled skill, creativity, and consistent brilliance. \\
    \hline
    Use of concessive claims & Continuing to make one's own arguments while agreeing with part of the other person's point ~\cite{tanskanen_concessive_2008} & I agree that any man who commits sexual assault is trash. But let's talk about Messi's personal life. \\
    \hline
    Use of advocacy claims & Arguing that certain conditions should exist ~\cite{rottenberg_structure_2014} & First of all, comparing team honors \textbf{should} not be done in terms of numbers alone, but rather in terms of the gold content of the championships as well as the degree of contribution to the tournament as a whole. \\
    \bottomrule
  \end{tabular}
\end{table*}

\begin{table*}[h!]
  \caption{Emerging "evidence" patterns from the forum.}
  \begin{tabular}{>{\raggedright\arraybackslash}p{2.5cm}>{\raggedright\arraybackslash}p{4cm}>{\raggedright\arraybackslash}p{9.5cm}}
    \toprule
    \textbf{Patterns} & \textbf{Description} & \textbf{Examples}\\
    \midrule
    Use of statistics (numerical data) in the evidence & Using numerical summaries ~\cite{rieke_argumentation_2012} & Moreover, Messi has provides more assists \textbf{(318 vs. 229)} than Ronaldo. And his goal-to-game ratio is superior, averaging \textbf{0.87} goals per game compared to Ronaldo's \textbf{0.77}. \\
    \hline
    Use of background information in the evidence & Providing contexts that are essential for understanding the arguments ~\cite{lunsford_everythings_2018} & \textbf{Even when injured in the 2016 Euro final}, Ronaldo led and inspired his teammates from the sidelines, playing a key role in Portugal's triumph. \\
    \hline
    Use of personal observations in the evidence & Justifying claims based on what has been directly seen or experienced ~\cite{kuhn_students_2005} & I would like to bring up the best argument for Messi being the GOAT, the recent world cup win. \textbf{I will admit I thank my lucky stars that I was able to witness a football match like that.} \\
    \hline
    Use of facts in the evidence & Providing evidence that the audience will accept as as being objectively verifiable ~\cite{rottenberg_structure_2014} & Ronaldo has faced serious allegations of sexual assault. \\
    \hline
    Use of examples in the evidence & Using instances to provide the empirical grounding for the claims~\cite{rieke_argumentation_2012} & The most impressive examples are his hat-trick against Spain in the 2018 World Cup and his hat-tricks against Wolfsburg and Atletico Madrid in the Champions League. \\
    \hline
    Use of counterexamples in the evidence & Providing a possibility that is consistent with the premises but inconsistent with the conclusion ~\cite{johnson-laird_how_2008} & \textbf{Alice}: ... In my opinion, the GOAT should excel under different coaches and with different teammates. Messi found a system that worked for him at Barcelona and stayed, but it's hard to argue he'd be just as successful at other clubs. The ability to thrive in various environments is crucial, and that's where Ronaldo has proven himself superior. \textbf{Bob}: Playing for more clubs and countries should not be considered as better. \textbf{On the other hand Ronaldo did not show dominance after his first Man Utd and Real years.} \\
    \bottomrule
  \end{tabular}
\end{table*}

\begin{table*}[h!]
  \caption{Emerging "reasoning" patterns from the forum.}
  \begin{tabular}{>{\raggedright\arraybackslash}p{2.5cm}>{\raggedright\arraybackslash}p{4cm}>{\raggedright\arraybackslash}p{9.5cm}}
    \toprule
    \textbf{Patterns} & \textbf{Description} & \textbf{Examples}\\
    \midrule
    Use of rebuttal in reasoning & Countering the validity of previous arguments by setting aside the general authority of warrant ~\cite{toulmin_uses_2003} & "Messi is not the most important player in the final" means that you did not watch the game. Football is not only about scoring. \\
    \hline
    Use of analogy in reasoning & Making a comparison between two similar cases and inferring that what is true in one case is true in the other ~\cite{freeley_argumentation_2008} & Football is not only about scoring. \textbf{If so, Anthony is probably better than Modric.}\\
    \hline
    Use of irrelevant conclusion as fallacies in reasoning & Ending with a conclusion that is not related in any necessary way to the premises ~\cite{kord_grey_2021} & Market value means nothing lol, Messi is 2 years younger than Ronaldo and that's why.\\
    \hline
    Use of hasty generalization as fallacies in reasoning & Trying to support a general claim by offering a story, which is just a single incident ~\cite{kord_grey_2021, van_eemeren_argumentation_2016} & Despite Messi and Ronaldo they have both won the Ballon d'Or and their individual abilities are outstanding, Messi is the one who led the Argentine national team to win the World Cup and Copa America, which has fully demonstrated his leadership ability. Since 2022 was a tough time for Argentina, Messi's achievement of leading Argentina to the World Cup somehow gave the Argentine people a great encouragement. That's why I consider Messi is more outstanding than Ronaldo, cause he knows how to be a better leader and how to cheered his team up. \\
    \hline
    Use of ad hominem attack as fallacies in reasoning & Rejecting or dismissing another person's statement by attacking the person rather than the statement itself ~\cite{kord_grey_2021, van_eemeren_argumentation_2016}& Chris [Carol] you have 0 ball knowledge. \\
    \hline
    Use of misplacing the burden of proof as fallacies in reasoning & Arguing that something is true simply because no one has proved it false, or that something is false simply because no one has proved it true ~\cite{kord_grey_2021} & I recognize that Ronaldo has not won a World Cup. Soccer is a team sport and Portugal is a little less strong overall. How many of Argentina's goals to win the World Cup came from Messi? How many of the goals were sporting goals? And how many goals came from penalties? \\
    \hline
    Use of straw man argument as fallacies in reasoning & Distorting or misrepresenting the opponent's argument, thus making it easier to knock it down or refute it ~\cite{van_eemeren_argumentation_2016} & Bob: In terms of team honor, whatever in National team or club team, Messi had more champion cup than CR7, with a total number of 45, and also has a biggest honor-- the FIFA world cup. In terms of personal honor, Messi has eight Ballon d' Or awards and more European Golden Shoe and some other personal honor. All in all, whether it is individual honor or team honor, Messi is better than CR7, CR7 outstanding place only more Champions League. \textbf{Alice: First of all, comparing team honors should not be done in terms of numbers alone, but rather in terms of the gold content of the championships as well as the degree of contribution to the tournament as a whole.} \\
    \bottomrule
  \end{tabular}
\end{table*}

\clearpage
\section{Number of Posts}
\label{C}
\begin{table*}[h!]
  \caption{An overview of post counts in the study.}
  \begin{tabular}{c|c|c|cc|c|ccc}
    \toprule
    \textbf{Session} & \textbf{Total} & \textbf{Part 1} &\textbf{Alice} &\textbf{Bob} &\textbf{Part 2} &\textbf{Alice} &\textbf{Bob} &\textbf{Carol}\\
    \midrule
    \textbf{I} & \textbf{18} & \textbf{6} & 3 & 3 & \textbf{12} & 3 & 5 & 4\\
    \textbf{II} & \textbf{15} & \textbf{6} & 3 & 3 & \textbf{9} & 3 & 3 & 3\\
    \textbf{III} & \textbf{20} & \textbf{4} & 2 & 2 & \textbf{16} & 7 & 6 & 3\\
    \textbf{IV} & \textbf{31} & \textbf{6} & 3 & 3 & \textbf{25} & 11 & 9 & 5\\
    \textbf{V} & \textbf{18} & \textbf{6} & 3 & 3 & \textbf{12} & 6 & 3 & 3\\
    \textbf{VI} & \textbf{18} & \textbf{6} & 3 & 3 & \textbf{12} & 6 & 3 & 3\\
    \textbf{VII} & \textbf{17} & \textbf{6} & 3 & 3 & \textbf{11} & 5 & 3 & 3\\
    \textbf{VIII} & \textbf{14} & \textbf{4} & 2 & 2 & \textbf{10} & 3 & 4 & 3\\
    \textbf{IX} & \textbf{20} & \textbf{4} & 2 & 2 & \textbf{16} & 10 & 3 & 3\\
    \textbf{X} & \textbf{13} & \textbf{4} & 2 & 2 & \textbf{9} & 3 & 3 & 3\\
    \textbf{XI} & \textbf{15} & \textbf{4} & 2 & 2 & \textbf{11} & 5 & 3 & 3\\
    \textbf{XII} & \textbf{13} & \textbf{4} & 2 & 2 & \textbf{9} & 3 & 3 & 3\\
    \textbf{XIII} & \textbf{44} & \textbf{6} & 3 & 3 & \textbf{38} & 35 & 3 & 0\\
    \bottomrule
  \end{tabular}
\end{table*}

\end{document}